\documentclass[twocolumn,preprintnumbers,amsmath,amssymb,A4]{revtex4}

\usepackage{graphicx}
\usepackage{dcolumn}
\usepackage{bm}

\begin{document}

\title{Hf defects in HfO$_{2}$/Si}

\author{W. L. Scopel}
\email{wlscopel@if.uff.br}
\address{Universidade Federal Fluminense, Escola de Engenharia Industrial 
Metal\'urgica de Volta Redonda,
Av. dos Trabalhadores, 420, CEP 27255-250, Volta Redonda, RJ, Brazil}
\author{Ant\^onio J.R. da Silva}
\author{A. Fazzio}
\address{Instituto de F\'{\i}sica, Universidade de S\~ao Paulo, Caixa Postal 
66318, CEP 05315-970, S\~ao Paulo, SP, Brazil}

\date{\today}

\begin{abstract}

We investigate the possibility that Hf defects exist
in the Si channel of HfO$_2$/Si-based metal-oxide-semiconductor
devices. We have studied, using \textit{ab initio} Density Functional Theory
calculations, substitutional and interstitial Hf impurities in
\textit{c}-Si, for various charge states. Our results indicate that 1) the 
tetrahedral interstitial defect is energetically more favorable than the
substitutional, and 2) there are various stable charge states in the Si gap.
The possible presence of these charged impurities in the Si channel
could lead to a mobility reduction, due to coulombic scattering.

\end{abstract}

\maketitle

The continued scaling down of silicon-based complementary
metal-oxide-semiconductor (CMOS) technology has required
intense search for high-k gate dielectric materials that can
replace SiO$_2$. In particular, HfO$_2$ has been considered
as one of the most promising candidates, due to its large
band gap, high dielectric constant, low leakage current, and
its thermodynamic stability on Si \cite{hobbs,shiraishi}.
However, for the sucessful integration of this high-k
material with the Si/CMOS technology there are still a number
of fundamental issues that must be solved, such as reduced
channel mobility and charge trapping states at the oxide.

Previous works\cite{torii,casse} have reported that the channel mobilities in CMOS
devices with HfO$_2$ gate dielectrics are significantly degraded when compared to
SiO$_2$. A possible mechanism to explain this behavior is the remote Coulomb
scattering (RCS) due to charge trapping centers at the Si/HfO$_2$ interface.
We here investigate, using state-of-the-art first principles calculations, possible
additional charge trapping centers related to Hf impurity defects in Si. 
Even though under equilibrium conditions Hf
atoms do not have a significant solubility in Si,
non-equilibrium growth conditions where the HfO$_2$ is
grown directly on top of the Si substrate may lead to the
incorporation of a small number of Hf atoms as impurities,
in a narrow region close to the Si/HfO$_2$ interface
\cite{lopez,renault}.

\begin{figure}[h]
\includegraphics[width=8.0cm]{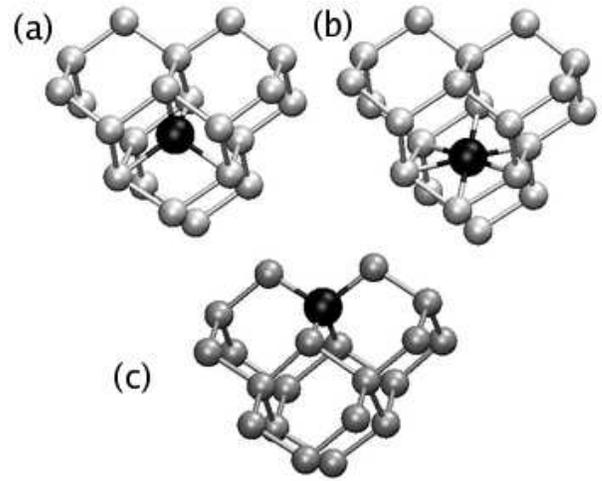}
\caption{\label{geometry}
Schematic geometries of defects in $c$-Si. Atomic structure of (a) a tetrahedral
interstitial, (b) a hexagonal site, and (c) a substitutional defect.
The black spheres (larger) represent Hf atoms, and the gray spheres (smaller)
represent Si atoms.}
\end{figure}

In this work, we have investigated the formation energy of
different Hf impurity defects in \textit{c}-Si,
through first-principles calculations, based on
the Density Functional Theory (DFT). In particular, we study
Hf at substitutional (Hf$_{Si}$) and interstitial sites.
For interstitial Hf defects, we have studied two different
configurations: (i) Hf at tetrahedral sites (Hf$_I^T$); and (ii) Hf at
hexagonal sites (Hf$_I^H$). For all cases studied we considered
various charges states $q$. The main conclusion is that there
are many charge states in the Si gap associated with these
defects, which may lead also to detrimental coulombic scattering.

The DFT calculations were performed using ultrasoft Vanderbilt
pseudopotentials \cite{Vanderbilt}, and the generalized gradient
approximation (GGA) for the exchange-correlation potential as
implemented in VASP code \cite{Perdew,vasp1,vasp2,vasp3}. In order to study the
interstitial and substitutional defects in Si, we have used 129 atoms
(128 Si atoms and 1 Hf atom) and 128 atoms
(127 Si atoms and 1 Hf atom) supercells, respectively.
We have used a plane-wave-cutoff energy of 151 eV
and the Brillouin zone was sampled at the $L$-points.
In all calculations the atoms were allowed to relax until the
atomic forces were smaller than 0.025 eV/\AA. Spin polarized
calculations were also performed, but we observed that the effect were very
small regarding the total energies. Thus, we will report below only the non-spin polarized
results.

\begin{figure}[h]
\includegraphics[width=8.3cm]{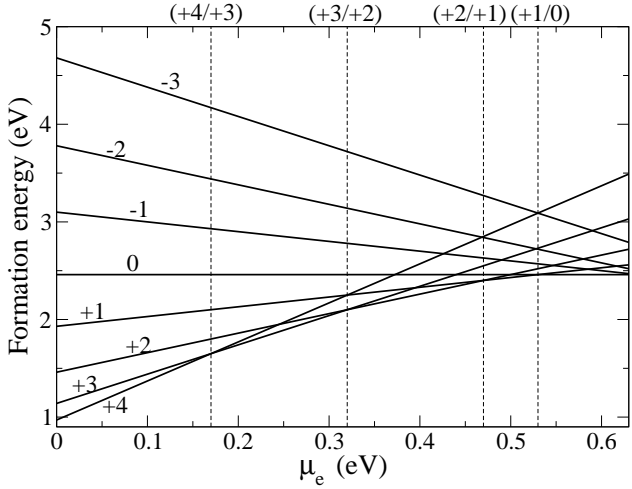}
\caption{\label{energy_tetr} Formation energy as function of the fermi level
$\mu_e$ for Hf defects at tetrahedral interstitial sites in \textit{c}-Si,
for a hafnium -rich growth condition. The numbers on the lines
indicate defects charge states.}
\end{figure}

The final relaxed geometries for the neutral defects are presented
in Fig. \ref{geometry}. In Fig. \ref{geometry}(a) we show the tetrahedral
interstitial defect. The Hf has four Si nearest neighbors with Hf-Si bonds of
approximately 2.50~\AA. In Fig. \ref{geometry}(b) the Hf$_I^H$ defect is presented. The Hf
atom has now six Si neares neighbors with average Hf-Si bond lengths of
2.54~\AA. Finally, in Fig. \ref{geometry}(c) the Hf$_{Si}$ is shown, and
as can be seen the Hf atom has also four Si nearest neighbors with Hf-Si
bondlengths of approximtely 2.57~\AA. Therefore, there is a global outward
relaxation of about 14\% for the Hf-Si bonds when compared to the
original Si-Si bonds, which is caused by the larger Hf atomic radius. We have also investigated
the possible existance of a dumbbell structure for the intertitial defect, since
this is the lowest energy configuration for the Si self-interstitial.
However, this structure turned out to be unstable, relaxing towards the
tetrahedral defect. This is most liked caused by the larger Hf radius. For all the
different charge states the final geometries were only slightly modified, with
variations in the Hf-Si bonds of less than 0.05~\AA. Small Jahn-Tell distortions
were also observed in some cases. However, none of these effects alter in
any qualitative way our main conclusion, which is the existence of charge states
in the gap. Therefore, we will not analize them in any further detail below.

The formation energy for a substitutional defect in the charge
state $q$ is calculated as
\begin{eqnarray}
E_{f}^{q}(Hf_{Si}) = E_{t}^{q}(Hf_{Si})-\frac{N_{Si}}{N}E_{t}(c\text{-Si})\nonumber\\
-\mu_{Hf}+q(\mu_{e}+E_v),
\label{subs}
\end{eqnarray}
\noindent whereas for the interstitial defects in the charge state $q$ it is calculated as
\begin{eqnarray}
E_{f}^{q}(Hf_I) = E_{t}^{q}(Hf_I)-E_{t}(c\text{-Si})\nonumber\\
-\mu_{Hf}+q(\mu_{e}+E_v)
\label{inter}
\end{eqnarray}

\begin{figure}[h]
\includegraphics[width=8.3cm]{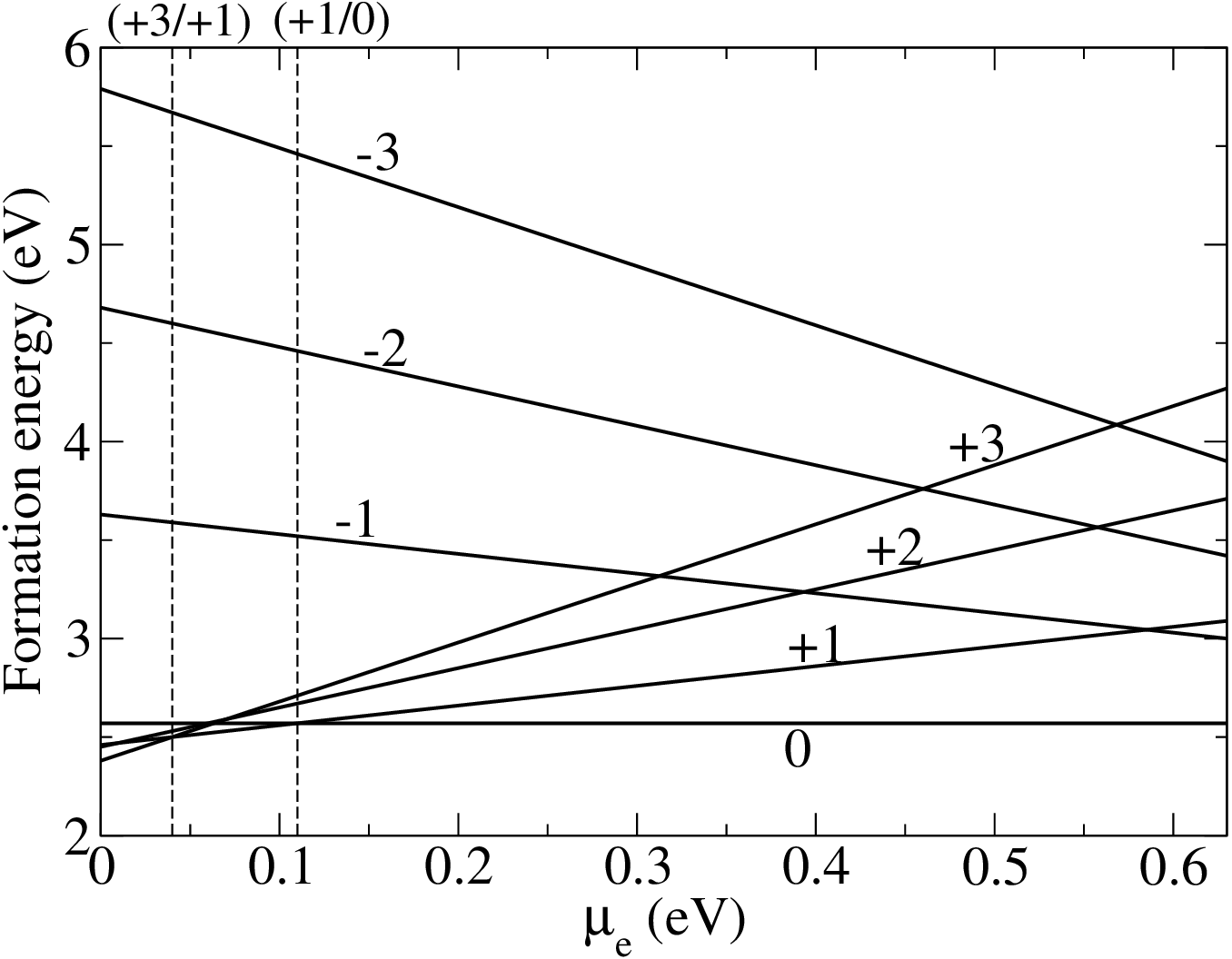}
\caption{\label{substitutional}Formation energy as
function of the fermi level $\mu_e$ for Hf defects
at substitutional interstitial sites in \textit{c}-Si for a Hf-rich growth condition.
The numbers on the lines indicate defects charge
states.}
\end{figure}

In the above expressions, E$^{q}$$_{t}$($D$) are the total
energies of the fully relaxed supercells with the
substitutional or interstitial defect $D$, and E$_{t}$(c-$Si$)
is the total energy of the similar supercell for the perfect
crystal of \textit{c}-Si; $\mu_{e}$ is electronic chemical potential,
$N_{Si}$ is the number of Si atoms and $N$ is
the total number of atoms in the supercell. We have considered the
bulk c-$Si$ as the source of Si atoms. The value for
the Hf chemical potential, $\mu_{Hf}$, depends on the growth conditions.
We will consider the limits of Hf-rich conditions and O-rich conditions
similarly to our previous work \cite{scopel}.

For the neutral interstitials, we obtain that the formation energy
for the Hf$_I^H$ is 2.3 eV larger than for the Hf$_I^T$. Therefore,
we will only analyse in detail below the Hf$_I^T$ defect.
In Fig.~\ref{energy_tetr} we present the formation energies for the
Hf$_I^T$ defect in different charge states $q$, calculated via
Eq.~\ref{inter}. The formation energies are plotted as a function of
$\mu_e$, which is varied between the theoretical values of
the valence band maximum ($\mu_e=0$ eV) and conduction band minimum
($\mu_e=0.63$ eV). The values reported in Fig.~\ref{energy_tetr} are
for Hf-rich conditions \cite{scopel}. For oxygen rich conditions the
curves would have to be up shifted by 10.9 eV. We have studied
charge states with $q$ varying from (+4) to (-3). As can be seen,
there are many equilibrium charge states in the gap, five total, from
(+4) to (0). This gives rise to four transition levels, the
(+4/+3), (+3/+2), (+2/+1), and (+1/0) at $\mu_e$ equal to 0.17 eV,
0.32 eV, 0.47 eV, and 0.53 eV, respectively. This indicates that for
a large range of electronic chemical potentials there will be charged
states that can contribute to coulombic scatering at the channel.
It is importatant to stress that the particular value of the transition
level may change due to the well known limitation of DFT in describing
the band gap (our theoretical band gap is 0.63 eV whereas the experimental
gap is 1.17 eV). However, the important point of our paper, which is
the existence of these charged states, does not depend on these possible
shifts of the transition levels.

\begin{figure}[h]
\includegraphics[width=8.3cm]{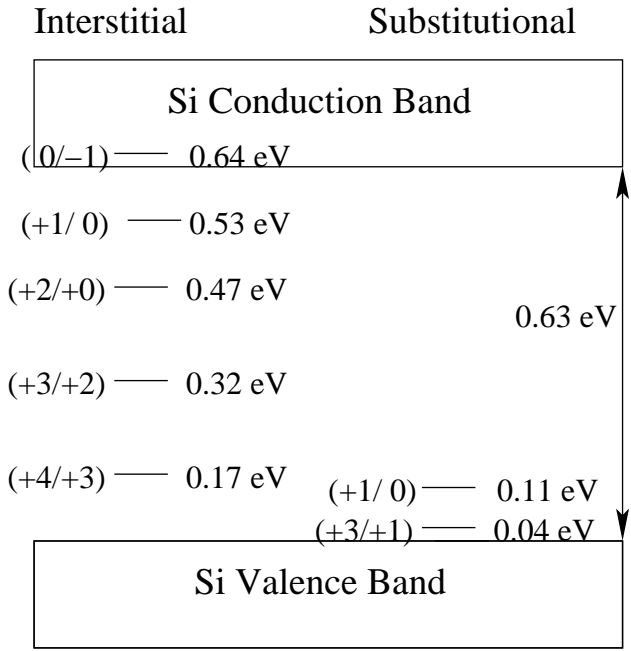}
\caption{\label{niveis} Schematic structure for the transition levels in the gap,
for the Hf$_I^T$ impurity defect (left) and the Hf$_{Si}$ defect (right). We use
the theoretical band gap of 0.63 eV.}
\end{figure}

The formation energies for substitutional Hf defects, calculated
according to Eq.~\ref{subs}, are presented 
in Fig.~\ref{substitutional}. The choices of chemical potentials
are the same as in Fig.~\ref{energy_tetr}. As can be seen,
the neutral defect is stable for a large range of the electronic
chemical potential. This is expected due to the ''isovalent character''
of Hf when compared to Si. However, for $\mu_e$ close to the
top of the valence band there are two transition levels. Very close
to the top of the valence band the (+3) charge
state is stable, with the (+3/+1) transition level at $\mu_e=0.04$ eV,
having, thus, a small negative-U character.
The second (+1/0) transition level is at $\mu_e=0.11$ eV.

Comparing the interstitial and substitutional defects, we see that
the Hf$_I^T$ is more stable than the Hf$_{Si}$ in the entire range of the
electronic chemical potential. For p-type materials the Hf$_I^T$ is almost
1.5 eV more stable than the Hf$_{Si}$, whereas for n-type materials
this same difference is only of the prder of 0.1 eV.

In summary, as shown in Fig. \ref{niveis}, we have shown that Hf impurity defects in Si will lead to
many stable charged states in the gap, resulting in detrimental
coulombic scattering and reducing the mobility in the channel.
For p-type Si, the interstitial Hf impurities will be in a (+4)
charge state, and they are more stable than the substitutional sites
by more than 1.25 eV. However, if only substitutional impurities
were present due to the growth conditions, they would also be charged.
For n-type material the neutral substitutional defect is now more stable,
but with formation energies very close to ones for the interstitial defect.
In any case, it seems that Hf may be more detrimental in p-type Si.
Finally, it would be very important if experimental works could investigate
the possible presence of these impurities.

\begin{acknowledgments}

Thanks are due to the Brazilian agency FAPESP and CNPQ for financial support.
We also thank CENAPAD-SP (Brazil) for the computer time.

\end{acknowledgments}

\end{document}